\documentclass[letterpaper]{article} 
\usepackage{aaai25}  
\usepackage{times}  
\usepackage{helvet}  
\usepackage{courier}  
\usepackage[hyphens]{url}  
\usepackage{graphicx} 
\usepackage{natbib}  
\usepackage{caption} 
\usepackage{algorithm}
\usepackage{algorithmic}
\usepackage{amsthm}

\usepackage{newfloat}
\usepackage{listings}
\DeclareCaptionStyle{ruled}{labelfont=normalfont,labelsep=colon,strut=off} 
\floatstyle{ruled}
\newfloat{listing}{tb}{lst}{}
\floatname{listing}{Listing}
\title{REM: A Scalable Reinforced Multi-Expert Framework for Multiplex Influence Maximization}
\author{
    Huyen Nguyen  \equalcontrib \textsuperscript{\rm 1},
    Hieu Dam \equalcontrib \textsuperscript{\rm 2},
    Nguyen Do \textsuperscript{\rm 3},
    Cong Tran \textsuperscript{\rm 1},
    Cuong Pham \textsuperscript{\rm 1}
}
\affiliations{
    \textsuperscript{\rm 1}
    Posts and Telecommunications Institute of Technology, Hanoi, Vietnam \\
    \textsuperscript{\rm 2}
   FPT University, Swinburne Vietnam Hanoi campus, Hanoi Vietnam \\
    \textsuperscript{\rm 3}
    University of Florida, Gainesville, USA \\
    huyenxam250400@gmail.com, hieudmswh00475@fpt.edu.vn,   nguyen.do@ufl.edu,
    cuongpv@ptit.edu.vn,
    congtt@ptit.edu.vn,
    
}

\usepackage{bibentry}
\usepackage{graphicx}
\usepackage{caption}
\usepackage{algorithm}
\usepackage{algorithmic}
\usepackage{amsmath}
\usepackage{amssymb}
\usepackage{mathtools}
\usepackage{booktabs}
\usepackage{subfig}
\usepackage{mathtools}
\usepackage{float}
\usepackage{microtype}
\usepackage{multirow}
\usepackage{soul}
\usepackage{mathtools}
\usepackage{graphicx}
\usepackage{float}
\usepackage{lipsum}
\usepackage{subfig}
\usepackage{textcomp}
\usepackage{xcolor}
\usepackage{placeins}
\usepackage{hhline}
\usepackage{tabularx}
\usepackage{enumitem}
\usepackage{adjustbox}

\newtheorem{dfn}{\textbf{Definition}}
\newtheorem{thm}{Theorem}
\newtheorem{lem}[thm]{Lemma}

\begin{document}

\maketitle

\begin{abstract}
In social online platforms, identifying influential seed users to maximize influence spread is a crucial as it can greatly diminish the cost and efforts required for information dissemination. While effective, traditional methods for Multiplex Influence Maximization (MIM) have reached their performance limits, prompting the emergence of learning-based approaches. These novel methods aim for better generalization and scalability for more sizable graphs but face significant challenges, such as (1) inability to handle unknown diffusion patterns and (2) reliance on high-quality training samples. To address these issues, we propose the Reinforced Expert Maximization framework (REM). REM leverages a Propagation Mixture of Experts technique to encode dynamic propagation of large multiplex networks effectively in order to generate enhanced influence propagation. Noticeably, REM treats a generative model as a policy to autonomously generate different seed sets and learn how to improve them from a Reinforcement Learning perspective. Extensive experiments on several real-world datasets demonstrate that REM  surpasses state-of-the-art methods in terms of influence spread, scalability, and inference time in influence maximization tasks. 
\end{abstract}

%

\section{Introduction}

Graph data has found a wide range of applications such as social networks and data mining \cite{lim2015cross, liaghat2013application, nettleton2013data, bonchi2011influence, ngo2024charme}. One popular application is Influence Maximization (IM), which aims to identify a set of individuals that can maximize the spread of influence in a social network under a specific diffusion model. This problem is known to be NP-hard and has been extensively studied in various domains such as viral marketing \cite{domingos2001mining, kempe2003maximizing}. With the diversification of social platforms, many users on Online Social Networks (OSNs) like Facebook and Twitter are linking their accounts across multiple platforms.  These interconnected OSNs with overlapping users are referred to as Multiplex Networks. The structure of multiplex networks allows users to post information across various OSNs simultaneously, presenting significant value for marketing campaigns \cite{vikatos2020marketing, zhang2022interaction, jalili2017link}. 

The inner information propagation models on each OSN can vary, leading to differences in how information spreads and influences users across platforms. Consequently, it becomes crucial to customize influence maximization strategies that effectively exert influence over multiple platforms. This is known as Multiplex Influence Maximization (MIM). To date, Combinatorial Optimization (CO) algorithms for MIM \citep{zhan2015influence,zhang2016least,kuhnle2018multiplex,singh2019mim2, ling2023deep} have limitations compared to learning-based approaches. CO algorithms struggle to generalize to unseen graphs and handle diverse multiplex networks. They also face scalability issues when dealing with large-scale networks. Furthermore, CO algorithms rely on predefined rules or heuristics, limiting their ability to capture complex patterns and non-linear dependencies in multiplex networks. These shortcomings significantly undermine their effectiveness in optimizing the selection of influential seed nodes. In contrast, learning-based approaches \cite{do2024mimreasoner, yuan2024graph, chen2022touplegdd, li2018deeper} offer advantages in terms of generalization, scalability and capturing complex patterns. However, they still suffer critical weaknesses in MIM as follows: 

\begin{itemize}
    \item[1)] \emph{Inefficient optimization.} MIM, being a NP-hard problem with layers potentially scaling to billions, demands efficient training. RL methods, such as those presented in
    \cite{manchanda2020gcomb, chen2022touplegdd, yuan2024graph, do2024mimreasoner}, optimize seed sets in discrete spaces through exploration, iteratively improving solutions without an initial dataset. However, they rely on extensive random sampling, leading to long training time and risks of local optima. Data-driven approaches like \cite{ling2023deep} address these issues by leveraging generative models trained on diverse datasets, though their success is tied to dataset quality. Only with a sufficiently diverse training dataset can the model capture key features for optimization. Developing low-complexity models for efficient optimization remains a major challenge.

    \item[2)] \emph{Inaccurate propagation estimating models.} Accurately measuring propagation value is crucial for evaluating seed set effectiveness. Simulation-based methods \cite{manchanda2020gcomb, do2024mimreasoner, yuan2024graph} rely on running propagation processes to compute spread which is computationally expensive and scales poorly for large graphs. GNN-based approaches \cite{chen2022touplegdd, ling2023deep} predict spread more efficiently but face accuracy issues due to oversmoothing \cite{cai2020note}. This challenge is exacerbated in multiplex networks, where each layer may use a different propagation model and scale to billions of nodes, complicating accurate predictions.
    
\end{itemize}

\textbf{Our Contributions.} We propose Reinforced Expert Maximization (REM), a novel framework for tackling challenges in Multiplex Influence Maximization (MIM). First, we introduce Seed2Vec, a VAE-based model that maps the discrete, noisy input space into a cleaner, continuous latent space following a Gaussian distribution. This allows us to optimize the seed set within this latent space. To address Seed2Vec's reliance on training data quality, we frame it as a Reinforcement Learning (RL) policy, enabling efficient exploration of latent regions to generate novel seed sets with significant spread in multiplex networks. These samples are then used to iteratively retrain Seed2Vec, improving its performance. Finally, REM enhances spread estimation with the Propagation Mixture of Experts (PMoE), a method that employs multiple Graph Neural Network (GNN) models as experts to capture complex diffusion patterns. Experiments on real-world datasets show that REM outperforms state-of-the-art methods in influence spread, scalability, and inference time.

\section{Related Work}

\textbf{Combinatorial optimization for IM.} Influence Maximization is essentially a simplified instance of Multiplex Influence Maximization, constrained to a single network instead of encompassing multiple interconnected ones.  While traditional IM has witnessed significant advancements, MIM presents unique challenges due to the complex interplay between these interconnected layers. Early IM approaches relied heavily on simulation-based methods \cite{leskovec2007cost}, which involve repeatedly simulating the diffusion process on the network to estimate influence spread. These methods, while intuitive, can be computationally expensive, especially for large networks. Proxy-based methods \cite{kimura2006tractable, chen2010scalable_prevalent, chen2010scalable_lt} emerged to address scalability issues by approximating influence spread with simpler metrics. Leveraging the submodularity of influence diffusion, approximation algorithms like goyal2011celf++ \cite{goyal2011celf++} and UBLF \cite{zhou2015upper} provide efficient seed selection with guaranteed $(1-1/e)$-approximation ratios. Recently, Tiptop \cite{li2019tiptop} emerged as a game-changer, offering near-exact solutions to IM by achieving a $(1-\epsilon)$-optimal solution for any desired $\epsilon > 0$. Despite these advancements, MIM necessitates novel approaches due to the added complexity of multiple interconnected networks. While promising approaches utilizing combinatorial approximation algorithms \cite{zhang2016least} exist, MIM remains an active research area. Future directions include incorporating machine learning and leveraging specific multiplex network characteristics for more efficient and accurate solutions.

\textbf{Machine Learning for IM.} Learning-based methods, employing deep learning techniques, have emerged to overcome the limitations of traditional IM methods, particularly their lack of generalization ability. Integrating reinforcement learning (RL) with IM has shown potential \citep{lin2015learning, ali2018boosting}, with recent advancements focusing on learning latent embeddings of nodes or networks for seed node selection \citep{manchanda2020gcomb,chen2022touplegdd,li2022piano}. Graph neural networks (GNNs) have also been explored to encode social influence and guide node selection in IM \citep{ling2022source, yuan2024graph}. Specifically, DeepIM \citep{ling2023deep} leverages a generative approach for IM and achieves state-of-the-art performance. However, these methods are primarily designed for single-network IM problems and face challenges when extended to the more intricate MIM problem.  This is primarily due to the scalability limitations in handling multiple networks and the difficulty in effectively capturing complex inter- and intra-propagation relationships within a multiplex network. Moreover, these methods are often restricted by their dependence on observation samples. While recent work \citep{do2024mimreasoner} utilizes probabilistic graphical models to represent the influence propagation process in multidimensional networks combined with reinforcement learning to find optimal seed sets, it faces scalability issues due to its reliance on traditional methods in the initial stages. 

\textbf{Our Method:} REM addresses the key weaknesses of combinatorial optimization and ML-based methods in MIM by tackling inefficiencies in optimization, propagation accuracy, and scalability. REM leverages advances in machine learning with Seed2Vec, a VAE-based model that maps seed selection into a continuous latent space, enabling smoother and faster optimization compared to traditional combinatorial methods. To address the common ML limitation of relying on static observation training datasets, REM incorporates reinforcement learning to dynamically explore this latent space, iteratively refining seed sets embeddings. For accurate propagation estimation, REM employs Propagation Mixture of Experts (PMoE), a set of GNN-based models specialized in capturing diverse diffusion dynamics across multiplex networks, mitigating oversmoothing and enabling accurate predictions. These enhancements improve REM's influence spread, scalability, and inference speed, addressing critical limitations of existing methods.

\section{Problem Formulation}

A multiplex network with $l$ layers is represented by $\mathcal{G}=\left\{G_1=\left(V_1, E_1\right)\right.$, $\left.G_2=\left(V_2, E_2\right), \ldots, G_l=\left(V_l, E_l\right)\right\}$, where each element consists of a directed graph $G_i=\left(V_i, E_i\right)$. If a node exists in more than one layer, then this node is added to set the overlapping users of the multiplex $\mathcal{G}$. Without loss of generality, we consider each layer of the multiplex has a same number of nodes. Therefore, if a node $v \in G_i$ does not belong to $G_j$ $\left(i \neq j\right)$ we add this node to $G_j$ as an isolated node. Then for each node, interlayer edges are added to connect its adjacent interlayer copies across all the multiplex networks. Finally, we consider the set of all nodes of the multiplex network as $V=\bigcup_{i=1}^l V_i$. In this study, since we permit different layers of a multiplex to follow distinct models of influence propagation, it is essential to define a mathematical model for the propagation on network $\mathcal{G}$.

\begin{dfn}[Influence Spread]

Given a layer $G_i=\left(V, E_i\right)$, we define a seed set $S \subseteq V$. The function $\delta_i$ represents an influence propagation model within $G_i$, which maps from the power set of $V$ to the non-negative real numbers, $\delta_{i}: 2^V \rightarrow \mathbb{R}{\geq 0}$. The influence spread, defined as the expected number of nodes influenced by the seed set $S$, is denoted as $\delta_i(S)$ and is calculated as follows:
\begin{equation} \label{eq: dfn_propagation_im}
\delta_{i}(S) = \lim_{m \to \infty} \frac{1}{m} \sum_{j=1}^{m} |T_j|,
\end{equation}

where $T_j$ represents the final activated sets $T_j \subset V$ given a seed set $S$ at the $j$-th simulation step. The simulation continues until no more nodes are activated or until reaching the maximum number of Monte Carlo steps,  $m$. Increasing $m$ improves the accuracy of estimating influenced nodes. This method is applicable to most propagation models, including Independent Cascade (IC) and Linear Threshold (LT) models \citep{kempe2003maximizing}.

\end{dfn}

Next, let us define the overall influence propagation model $\delta$ in the multiplex network $\mathcal{G}$. Firstly, when an overlapping node $v$ is activated in one layer graph $G_{i}$, its corresponding interlayer copies in other layers also become activated in a deterministic manner. This phenomenon is known as ``overlapping activation" \cite{kuhnle2018multiplex, do2024mimreasoner}, and is visualized in Figure \ref{fig: multiplex}. Secondly, when quantifying the expected number of influenced nodes in the entire multiplex network $\mathcal{G}$, we consider the overlapping nodes as a single instance rather than counting them multiple times. This means that when counting the influenced nodes across all layers, we do not add up the duplicates resulting from overlapping activation. Thus, the overall influence $\delta(S)$ combines the independent influences from each layer while accounting for overlapping activations:

\begin{equation} \label{eq: dfn_propagation_mim}
\delta(S) = \lim_{m \to \infty} \frac{1}{m} \sum_{j=1}^{m} \left|\bigcup_{i=1}^l T_{ij}(S)\right|
\end{equation}

where $T_{ij} \subset V$ represents the final activated sets in layer $i$ given a seed set $S$ at the $j$-th simulation step. We are now ready to define our MIM problem as follows:

\begin{figure}[t]
\centering
\includegraphics[width=0.75\linewidth]{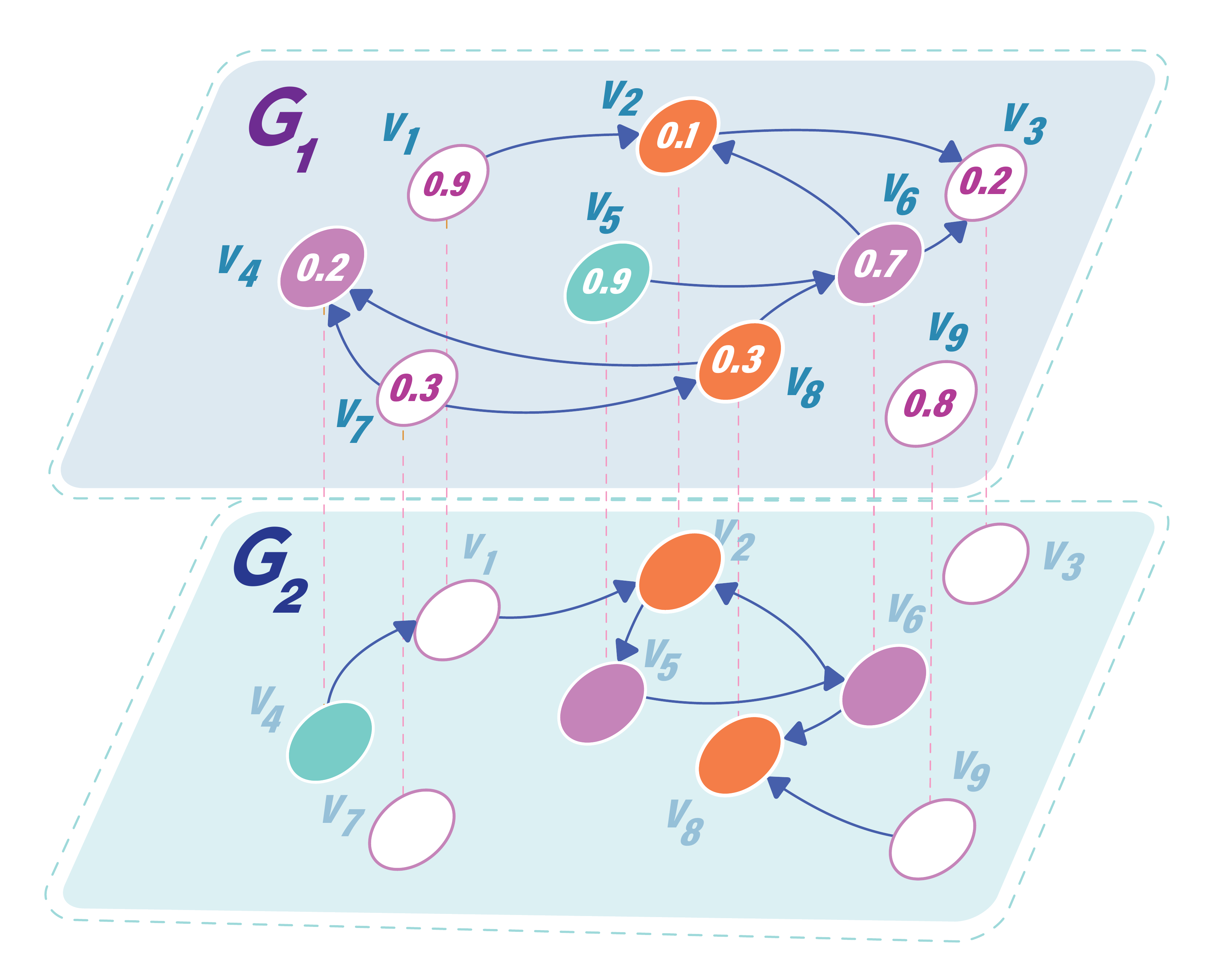}
   \caption{
   An example illustrating the unique "overlapping activation" property of influence propagation in a multiplex network. Two layers $G_1$ and $G_2$ has their own respective diffusion model, LT and IC. Orange nodes represent seed nodes, pink nodes are infected, and green nodes are nodes that are activated due to overlap. If propagation occurs independently within each layer, node $v_6$ of $G_1$ is inactive due to its high threshold of $0.7$, meaning it requires at least $70\%$ activated neighbors to become activated. However, in the multiplex, overlapping node $v_5$ of $G_1$ is also activated due to deterministic activation from $G_2$, meeting activation requirement of node $v_6$ in $G_1$. Therefore, total infected node set of the multiplex, not counting the same node in different layer, is now ($v_2$, $v_4$, $v_5$, $v_6$, $v_8$).
}
   \label{fig: multiplex}
\end{figure}

\begin{dfn}[Multiplex Influence Maximization (MIM)]
Given a multiplex graph $\mathcal{G} = {(G_1 = (V, E_1),\delta_1), \dots, (G_l = (V, E_l), \delta_l)}$ and a budget $b \in \mathbb{N}$. Specifically, seed set $S$ is represented as a binary vector  $\boldsymbol{x} \in \mathbb{R}^{1 \times |V|}$, where each element $\boldsymbol{x}_j$ corresponds to a node $v_j$ in $V$. Specifically, $\boldsymbol{x}_j = 1$ if $v_j$ is included in the seed set, and $\boldsymbol{x}_j = 0$ otherwise. Suppose we have a training dataset of seed set indicators pairs $(\boldsymbol{x}, y)$, where $\boldsymbol{x}$ represents a seed set and $y = \delta(\boldsymbol{x})$ is the corresponding total number of infected nodes. The MIM problem asks us to find an optimal seed node set $\tilde{\boldsymbol{x}}$ of size at most $b$ to maximize the overall influence spread $\delta(\boldsymbol{x})$ calculated in the multiplex. This problem is formulated as follows:

\begin{equation} \label{eq: multiplex}
\tilde{\boldsymbol{x}}=\arg \max _{|\boldsymbol{x}| \leq b} \delta \left(\boldsymbol{x}\right)
\end{equation}

\end{dfn}

For each layer $G_{i} \in \mathcal{G}$, many greedy based algorithms \citep{leskovec2007cost,goyal2011celf++,tang2014influence,tang2015influence} have obtained a performance guarantee bound of  $(1-1/e)$, if $\delta_{i}$ is submodular and monotone increasing \citep{kempe2003maximizing}. If all $\delta_{i}$ of all $G_{i}$ satisfy the Generalized Deterministic Submodular (GDS) property, then $\delta$ is submodular \citep{kuhnle2018multiplex}.

\section{Our Framework: REM}

\begin{figure*}[h]
\centering
\includegraphics[width=0.75\linewidth]{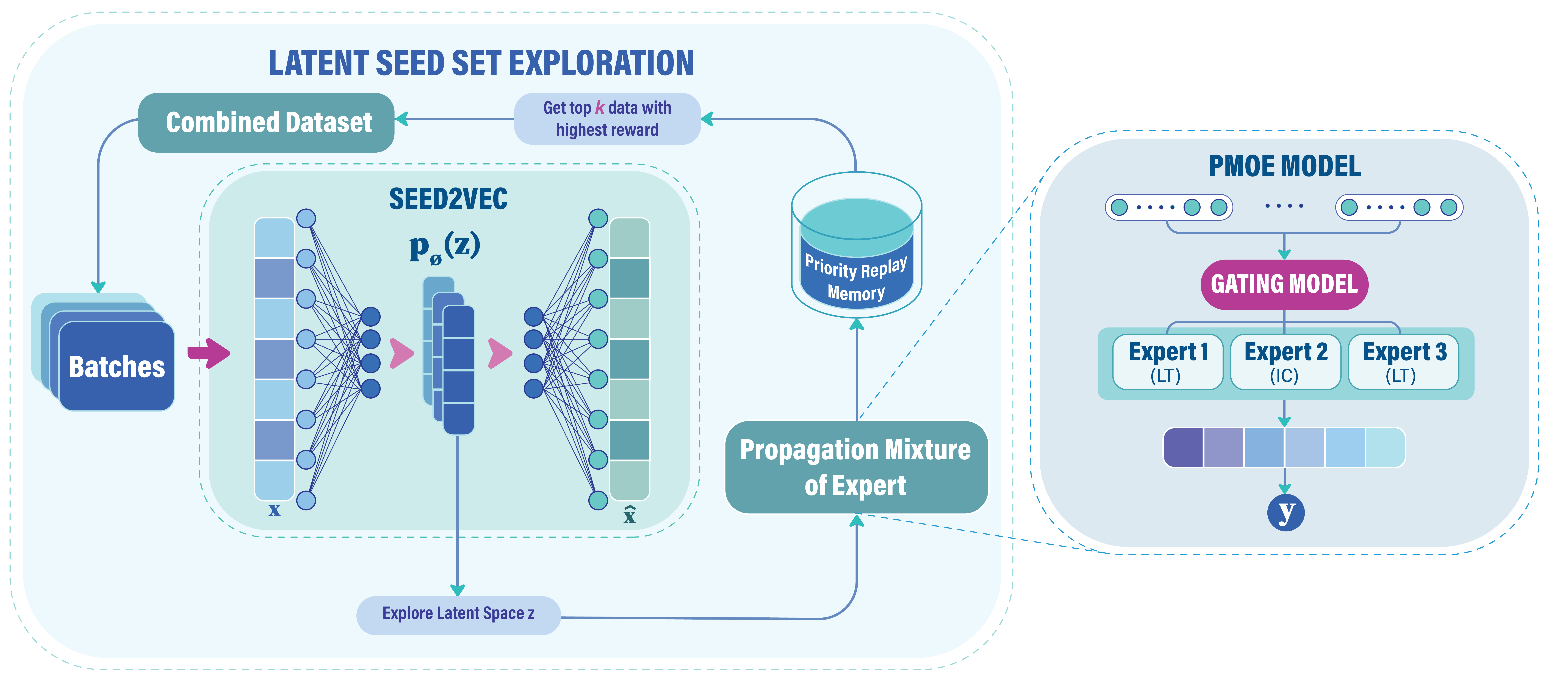}
\caption{The diagram depicts REM's process for addressing the MIM problem. Initially, REM utilizes Seed2Vec to embed complex representations of seed sets into a continuous and less noisy space. Subsequently, REM explores and generates various diverse seed sets from this latent space. REM maintains control over the quality of seed set generation through the Propagation Mixture of Experts (PMoE), a model capable of accurately learning and predicting the propagation of a given seed set in a large-scale multiplex network. Once the synthetic sets are generated, they are stored in a priority replay memory. To prevent model collapse or catastrophic forgetting, the top $k$ seed sets, which PMoE predicts to contribute the most to propagation in multiplex networks, are then combined with the original collected dataset to construct a new dataset for model retraining. This process reinforces the capability to produce higher-quality seed sets in future iterations. 
}
\label{fig: rem_framework}
\end{figure*}

The REM model addresses mentioned challenges by following concepts illustrated in Figure \ref{fig: rem_framework}. First, instead of optimizing the seed set in a complex and discrete space, REM employs our proposed Seed2Vec, a Variational Autoencoder (VAE)-based model \cite{kingma2013auto}. VAE is a generative framework that encodes data into a continuous latent space while preserving meaningful structure, enabling the representation of complex seed sets in a less noisy form. This allows for optimization and the generation of new potential solutions within that space. Recognizing that Seed2Vec only captures and generates solutions within the feature distribution of the original training data, our framework treats Seed2Vec as an RL agent. This agent explores and exploits diverse latent representations during each training episode. For each latent sample generated by Seed2Vec, we apply our proposed Propagation Mixture of Experts (PMoE) to predict its propagation with very high accuracy, rank, and store it in a Priority Replay Memory (PRM) \cite{horgan2018distributed} , a structure designed to prioritize important samples based on their predictive value for enhanced learning efficiency. We then sample the top $k$ samples from PRM and combine them with the original dataset to form a combined dataset. Finally, REM uses this combined dataset to retrain the Seed2Vec model, enhancing its seed set generation capability. More algorithmic details can be found in Appendix \textbf{A}.

\subsection{Seed2Vec: Learning To Embed Complex Seed Set}
To optimize and identify quality seed sets in a multiplex network, we propose characterizing the probability of a seed node set, denoted as $p_\phi(\boldsymbol{x})$, given the multiplex graph $\mathcal{G}$. Learning $p_\phi(\boldsymbol{x})$ can provide insights into the underlying nature of the seed set, facilitating effective exploration of seed sets in the search space. However, learning such a probability is challenging due to the interconnections between different nodes within each seed set and their high correlation based on the network topology of $\mathcal{G}$. These complex connections make the node relationships within seed sets difficult to decipher compared to other similar combinatorial problems. Therefore, instead of learning directly the complex representation of $\boldsymbol{x}$, we learn a latent presentation $\boldsymbol{z}$ using Variational Auto Encoder (VAE) \citep{kingma2013autoencoding} denoted as $\mathcal{F}_\theta$. For convenient, we further decompose the VAE model $\mathcal{F}_\theta$ into two models: the Encoder denoted as $\mathcal{E}_\psi$ and the Decoder model denoted as $\mathcal{D}_\phi$. Formally, we have:
\begin{equation} \label{eq: dfn_vae_model}
    \mathcal{F}_\theta=\mathcal{E}_\psi \circ \mathcal{D}_\phi, \quad \hat{\boldsymbol{x}}=\mathcal{F}_\theta\left(\boldsymbol{x}\right)=\mathcal{D}_\phi\left(\mathcal{E}_\psi\left(\boldsymbol{x}\right)\right)=\mathcal{D}_\phi(\boldsymbol{z}),
\end{equation}

where $\hat{\boldsymbol{x}}\in [0,1]^{1 \times |V|}$ represents the reconstructed seed set generated.

Specifically, to generate $\boldsymbol{x}$, $\mathcal{F}_{\theta}$ assumes the existence of a latent random variable $\boldsymbol{z} \in \mathbb{R}^{1 \times s}$, where $s$ represents the dimension of the variables in $\boldsymbol{z}$. This latent variable captures the features of the original seed set and follows a latent distribution $p_\phi(\boldsymbol{z})$. The complete generative process can be described by the equation:

\begin{equation}
p_\phi(\boldsymbol{z} \mid \boldsymbol{x})=\frac{p_\phi(\boldsymbol{x} \mid \boldsymbol{z}) p_\phi(\boldsymbol{z})}{p_\phi(\boldsymbol{x})}
\label{inferece VAE}
\end{equation}

However, computing the exact value of $p_\phi(\boldsymbol{x})=\int \ldots \int p_\phi(\boldsymbol{x}, \boldsymbol{z}) , d_{\boldsymbol{z}_0} \ldots d_{\boldsymbol{z}_v}$ is intractable, making the equation computationally challenging. To address this problem, $\mathcal{E}_\psi$ will learn $q_\psi$ which is approximated posterior distribution of $p_\phi(\boldsymbol{z} \mid \boldsymbol{x})$. The goal is to approximate the intractable posterior distribution with a simpler distribution $q_\psi(\boldsymbol{z} \mid \boldsymbol{x})$ given the seed set $\boldsymbol{x}$. In other words, the objective is to have $p_\phi(\boldsymbol{z} \mid \boldsymbol{x}) \approx q_\psi(\boldsymbol{z} \mid \boldsymbol{x})$.

This is used to derive the following Evidence Lower Bound (ELBO) to train the model using the reparameterization trick and SGD \citep{kingma2013autoencoding}.
\begin{equation}
\begin{aligned}
\mathcal{L}^{\text {ELBO }} & =\mathbb{E}_{q_\psi}\left[\log p_\phi(\boldsymbol{z}, \boldsymbol{x})\right] -\mathbb{E}_{q_\psi}\left[\log q_\psi(\boldsymbol{z} \mid \boldsymbol{x})\right] \\
& =\mathbb{E}_{q_\psi}\left[\log p_\phi(\boldsymbol{x} \mid \boldsymbol{z})\right]+\mathbb{E}_{q_\psi}\left[\log p_\phi(\boldsymbol{z})\right] \\
&
-\mathbb{E}_{q_\psi}\left[\log q_\psi(\boldsymbol{z} \mid \boldsymbol{x})\right] \\
& =\mathbb{E}_{q_\psi}\left[\log p_\phi(\boldsymbol{x} \mid \boldsymbol{z})\right]-\mathbb{E}_{q_\psi}\left[\log \frac{q_\psi(\boldsymbol{z} \mid \boldsymbol{x})}{p_\phi(\boldsymbol{z})}\right]
\end{aligned}
\label{eqn: loss_elbo}
\end{equation}

Note that we model $p_\phi(\boldsymbol{z})$ as a Gaussian distribution $\mathcal{N}(\mu, \sigma^2)$, where $\mu$ and $\sigma$ are defined hyperparameters. For a detailed decomposition of the ELBO and related details, refer to Appendix \textbf{B}.

\subsection{Propagation Mixture of Expert}

Applying Graph Neural Networks (GNNs) to predict propagation in large-scale multiplex networks with billions of nodes is challenging due to oversmoothing \cite{cai2020note}. In addition, when using a single GNN with $h$ layers, nodes aggregate information from $h$-hop neighbors, potentially mixing data from different layers, leading to inaccuracies. To overcome this, we propose the Propagation Mixture of Experts (PMoE). This approach uses multiple GNN models, each with different layer depths, to capture propagation dynamics effectively. Nodes are routed to the most suitable expert based on their characteristics and desired propagation depth, ensuring the model focuses on relevant information and reduces noise. This method allows accurate and efficient propagation prediction in large-scale multiplex networks.

Our PMoE framework captures the propagation process given a seed set $\boldsymbol{x}$ and a multiplex graph $\mathcal{G}$. In this framework, we define a set of $C$ "expert networks," denoted as $e_1, e_2, \dots, e_C$. Each expert $e_i$ is implemented as a GNN with varying layer depths, outputting $e_i(\boldsymbol{x}, \mathcal{G}, \xi_i) \in [0,1]^{1 \times |V|}$, a vector representing the estimated infection probability for each node in $\mathcal{G}$, where $\xi_i$ is the parameter of the $i$-th expert. To effectively leverage the diverse knowledge of experts, we employ a routing network $R$, which outputs a probability distribution over experts $R(\boldsymbol{x}) \in \mathbb{R}^{1 \times C}$ based on the input seed set $\boldsymbol{x}$. Each element in this distribution corresponds to the relevance probability of a particular expert for the given input. Inspired by the noisy top-$m$ routing mechanism proposed by \citep{shazeer2017outrageously}, we select the $m$ most relevant experts for each input. This mechanism operates as follows:
\begin{equation}
Q\left(\boldsymbol{x}\right)=\boldsymbol{x} \xi_g+\epsilon \cdot \operatorname{Softplus}\left(\boldsymbol{x} \xi_n\right),
\end{equation}
    \begin{equation}\label{eq: dfn_gating_function}
R\left(\boldsymbol{x}\right)=\operatorname{Softmax}\left(\operatorname{TopM}\left(Q\left(\boldsymbol{x}\right), m\right)\right),
\end{equation}

In this equation, $\epsilon \sim \mathcal{N}(0,1)$ represents standard Gaussian noise. The parameters $\xi_g$ and $\xi_n$ are learnable weights that control the contributions of the clean and noisy scores, respectively. The expected value $\mathcal{M}(\boldsymbol{x}, \mathcal{G} ; \xi)$, where $\xi = [\xi_g, \xi_n, \xi_1, \dots, \xi_C]$ represents the parameters of the PMoE model $\mathcal{M}$, is calculated based on the outputs of all experts and can be formulated as follows:

\begin{equation}\label{eq: dfn_pmoe}
\begin{aligned}
\mathcal{M}\left(\boldsymbol{x}, \mathcal{G} ; \xi\right) 
=\sum_{i=1}^C R_i(\boldsymbol{x}) e_i(\boldsymbol{x}, \mathcal{G}; \xi_i)
\end{aligned}
\end{equation}

Here, \(R_i(\boldsymbol{x})\) is the \(i\)-th element of routing network \(R(\boldsymbol{x})\), representing the relevance probability of the \(i\)-th expert in predicting the influence of seed set \(\boldsymbol{x}\). In this scenario, the total number of infected nodes, denoted as $\hat{y} \in \mathbb{R}_{+}$, is calculated as $\hat{y} = \mathcal{P}(\boldsymbol{x}, \mathcal{G} ; \xi)=g(\mathcal{M}\left(\boldsymbol{x}, \mathcal{G} ; \xi\right); \zeta)$. Here, $g(\cdot)$ is a normalization function (e.g., $l-1$ norm) and $\zeta$ is the threshold to transform the probability into discrete value.

\begin{lem}[\textbf{Monotonicity of PMoE Models}]
\label{lemma: monotonicity}
Assuming the PMoE model has been trained to convergence and during the inference phase, noisy scores $\xi_n$ are not considered, for any GNN-based, $\mathcal{P}$ is infection monotonic if the aggregation function and combine function in GNN are non-decreasing. (Proof in Appendix \textbf{C1})
\end{lem}

According to Lemma \ref{lemma: monotonicity} the PMoE model $\mathcal{P}(\boldsymbol{x}, \mathcal{G}; \xi)$ has the theoretical guarantee to retain monotonicity, and the objective of learning the PMoE model $\mathcal{P}(\boldsymbol{x}, \mathcal{G}; \xi)$ is given as maximizing the following probability with a constraint:
\begin{align}\label{eq: representation_y}
\max\nolimits_{\xi}\mathbb{E}\big[p_{\xi}(y|\boldsymbol{x}, \mathcal{G})\big], 
\end{align}

\subsection{Latent Seed Set Exploration}

As a generative model, Seed2Vec can only produce quality seed sets if the original training data is feature-rich. If the data is biased toward dominant features or lacks diversity, Seed2Vec may miss important but less prevalent features. As the multiplex becomes more complex and the number of nodes increases, the model tends to favor dominant seed nodes in the dataset, often overlooking less frequent but potentially significant ones. REM overcomes this by treating Seed2Vec as an RL agent, actively exploring novel and potentially impactful seed sets that maximize propagation to retrain and reinforce itself by the following lemma: 

\begin{lem}[\textbf{Latent Entropy Maximization Equivalence}]
\label{lemma: entropy}
Assuming the Seed2Vec model has convergened, we have $\arg \max _{\boldsymbol{z}} \mathcal{H}(\mathcal{D}_\phi(\boldsymbol{z}))
\propto \arg \max _{\boldsymbol{x}} \mathcal{H}(\boldsymbol{x})
$. (Proof in Appendix \textbf{C2})
\end{lem}

According to Lemma \ref{lemma: entropy}, exploration within the latent space $\boldsymbol{z}$, aimed at identifying the novel seed set $\mathcal{S}_t$, where $t=1, 2, 3, \dots$ represents the training episode, is proportional to exploration within the discrete space $\boldsymbol{x}$. This correlation emerges because a well-trained Seed2Vec model, using the original collected seed set $\boldsymbol{X}_0$, ensures both continuity—where nearby points in the latent space decode into similar content—and completeness, meaning that any point sampled from the latent space's chosen distribution generates 'meaningful' content. At this juncture, $p_\phi(\boldsymbol{z} \mid \boldsymbol{x}) \approx q_\psi(\boldsymbol{z} \mid \boldsymbol{x})$, with $q_\psi(\boldsymbol{z} \mid \boldsymbol{x})$ converging to a Gaussian distribution $\mathcal{N}(\mu, \sigma^2)$ as indicated by the second term of Equation \ref{eqn: loss_elbo}. Typically, an RL agent could explore various latent features by sampling $\boldsymbol{z} \sim \mathcal{N}(\mu, \sigma^2)$ and reconstructing the seed set $\hat{\boldsymbol{x}}$ using the Decoder (i.e., $\hat{\boldsymbol{x}} = \mathcal{D}_\phi(\boldsymbol{z})$). However, since $q_\psi(\boldsymbol{z} \mid \boldsymbol{x})$ converges to a continuous function that has a derivative with respect to $\boldsymbol{z}$. Instead of the RL agent exploring by random sampling, we use Gradient Descent directly on $\boldsymbol{z}$ to minimize the following objective function:

\begin{equation}
\mathcal{L}^{\mathrm{Explore}}( \boldsymbol{z}) = \mathbb{E} \left( c \cdot \mathcal{H}(\mathcal{D}_\phi(\boldsymbol{z})) +  \exp(-\mathcal{P}(\mathcal{D}_\phi(\boldsymbol{z})) \right)
\label{eqn: loss_explore}
\end{equation}

where $c$ are coefficients. The term $\mathcal{H}(\mathcal{D}_\phi(\boldsymbol{z}))=-\sum_{i=1}^{|\hat{\boldsymbol{x}}|} p(\hat{\boldsymbol{x}}_i) \log p(\hat{\boldsymbol{x}}_i)$ denotes the entropy of the latent variable, which promotes exploration within new regions of the latent space. The function $\mathcal{P}(\mathcal{D}_\phi(\boldsymbol{z}))$ refers to the Propagation Mixture of Experts (PMoE), which is detailed in the following section, and is used for predicting the reconstructed seed set $\hat{\boldsymbol{x}} = \mathcal{D}_\phi(\boldsymbol{z})$. To align with the objective of minimizing the loss function, we employ the exponential function, $\exp(\cdot)$, to reduce the impact of $\mathcal{P}(\mathcal{D}_\phi(\boldsymbol{z}))$ as its value increases. With the novel synthetic seed set $\mathcal{S}_t$ (store by using Priority Replay Memory \cite{schaul2015prioritized}) obtained by optimizing Equation \ref{eqn: loss_explore}, we sampling top $k$ best samples and combine them with the original dataset $\boldsymbol{X}_0$ to create a Combined Dataset $\boldsymbol{X}_t = \mathcal{S}_t^{(<k)} \cup \boldsymbol{X}_0$. Therefore, as training episode $t$ progresses, we can use $\boldsymbol{X}_t$ to retrain the Seed2Vec model $\mathcal{F}_\theta$. This approach allows $\mathcal{F}_\theta$ to generate improved seed sets in future iterations.

\noindent\textbf{End-to-end Learning Objective.} Finally, to bridge representation learning, latent seed set exploration, and diffusion model training, we minimize the following end-to-end objective function, which combines Eq. \eqref{eqn: loss_elbo}, \eqref{eqn: loss_explore}, and \eqref{eq: representation_y}:
\begin{align}\label{eq: learning_objective}
\mathcal{L}_{\text{train}}=\mathbb{E}\left[ \mathcal{L}^{ELBO}(\theta)+ \mathcal{L}^{PMoE}(\xi)+ \mathcal{L}^{Explore}( \boldsymbol{z})\right]
\end{align}
where $\mathcal{L}^{PMoE} = \left(\hat{y}-y\right)^2$.

\noindent\textbf{Seed Node Set Inference.} Finally, our method conclude with inferencing the seed node set from the continuous latent space. Specifically, gradient ascent is employed to find the latent representation $\tilde{\boldsymbol{z}}$ that maximizes the predicted influence spread, based on the estimation provided by the PMoE model. Representation $\tilde{\boldsymbol{z}}$ is decoded using the decoder network of Seed2Vec to obtain the optimal seed node set $\tilde{\boldsymbol{x}}$.

\begin{thm}[\textbf{Influence Estimation Consistency}]
\label{lemma: Consistency}
Given two distinct seed sets $\boldsymbol{x}^{(i)}$ and $\boldsymbol{x}^{(j)}$, with their corresponding latent representations $\boldsymbol{z}^{(i)}$ and $\boldsymbol{z}^{(j)}$ encoded by a Seed2Vec. 
If the reconstruction error is minimized during the training and $\mathcal{P}(p_{\phi}(\boldsymbol{z}^{(i)}), \mathcal{G}; \xi) > \mathcal{P}(p_{\phi}(\boldsymbol{z}^{(j)}), \mathcal{G}; \xi)$, then it follows that $\mathcal{P}(\boldsymbol{x}^{(i)},
\mathcal{G}; \xi) > \mathcal{P}(\boldsymbol{x}^{(j)}, \mathcal{G}; \xi)$. 
(Proof in Appendix \textbf{C3})
\end{thm}

According to Theorem \ref{lemma: Consistency} , the optimal seed set that maximizes influence can be found by optimizing $\boldsymbol{z}$.

\section{Experiment}

\begin{table*}[t]
    \centering
    \resizebox{\textwidth}{!}{%
    \begin{tabular}{@{}c|cccc|cccc|cccc|cccc|cccc|cccc|cccc@{}}
    \toprule
    &\multicolumn{4}{c|}{Cora-ML}
    & \multicolumn{4}{c|}{Celegans}      
    & \multicolumn{4}{c|}{Arabidopsis}   
    & \multicolumn{4}{c|}{NYClimateMarch2014} 
    &  \multicolumn{4}{c|}{ParisAttack2015}  \\ \midrule
    Methods & 1\%     & 5\%     & 10\%     & 20\%    & 1\%     & 5\%     & 10\%     & 20\%    & 1\%     & 5\%     & 10\%     & 20\%    & 1\%     & 5\%     & 10\%     & 20\%   & 1\%     & 5\%     & 10\%     & 20\%  
    \\ \midrule
    ISF          
    
    & \textbf{398.34}         & \textbf{778.62}         
    & \textbf{979.87}         & 1368.56 
    & \textbf{1465.86}         & \textbf{2298.01}         
    & 2571.92         & 2819.26          
    & 2415.04         & \textbf{3140.58}         
    & 3871.10         & 4694.16          
    & -              & -           
    & -              & -          
    & -              & -          
    & -              & -           
    \\
    KSN        
    & 398.31         & 778.62          
    & 979.10        & 1366.03  
    & 1382.86       & 2176.32          
    & 2335.44        & 2620.10         
    & 2282.72        & 2941.54         
    & 3621.26        & 4641.34         
    & -             & -           
    & -             & -          
    & -             & -          
    & -             & -            \\\midrule
    GCOMB       
    & 347.11          & 766.02          
    & 976.14          & 1251.52  
    & 1389.63         & 1896.86        
    & 2237.37         &2550.61         
    & 2315.44         & 3097.97
    & 3622.08         & 4547.63        
    & 2093.32            & 7228.02           
    & 11780.29           & 16933.89           
    & 114672.50            & 180977.09          
    & 356187.81           & 587891.16         
    \\
     ToupleGDD     
    & 349.01           &721.42            &862.42          &1132.66    
    & 1279.28         & 1905.23          
    & 2117.74         & 2411.14           
    & 2044.67         & 2856.37          
    & 3487.41         & 4483.27          
    & 1821.43            & 6714.98            
    & 10231.81           & 17822.21         
    & 102872.11              & 171992.43          
    & 335298.85              & 563387.05            
     \\
     DeepIM    
    &311.52           & 606.82             &826.41           &1179.45   
    & 1275.66         & 1527.25          
    & 1938.58         & 2251.53             
    & 1993.37        & 2397.89         
    & 3328.33         & 4073.50          
    & 1893.02            & 6409.32           
    & 8064.77           & 14269.49         
    & 83972.39            & 149237.59         
    & 281298.85           & 480393.42          
     \\
    MIM-Reasoner      
    & 398.22          & 778.02         
    & 978.95         & 1363.35     
    & 1432.39         & 2199.26         
    & 2389.91         & 2645.24          
    & 2396.95         & 2989.15          
    & 3729.39         & 4621.54         
    & 2101.86              & 7387.91          
    & 11984.55              & 21062.87         
    & 129650.48              & 217291.02           
    & 379932.17              & 608192.57         
    \\\midrule
    REM-NonRL       
    & 321.24          & 732.84          
    & 880.51          & 1151.68        
    & 1301.22         & 1884.95          
    & 2098.16         & 2451.87          
    & 2076.31         & 2421.13          
    & 3399.34         & 4120.51          
    & 2057.82            & 6835.44          
    & 8564.63           & 16021.63          
    & 87109.27              & 152621.23          
    & 298923.41              & 515237.59                \\
    
    REM-NonMixture      
    & 343.16          & 736.42          
    & 921.24          & 1301.55           & 1387.67         & 2062.79           & 2304.15         & 2695.28          
    & 2317.39         & 2982.67          
    & 3712.35         & 4611.27         
    & 2215.82            & 7124.29          
    & 10034.71           & 18932.40          
    & 125098.55              & 207621.48           
    & 380274.03              & 605875.67              
    \\
    \textbf{REM}     
    & 347.34 
    & 765.48          
    & 965.04 
    & \textbf{1404.14}  
    & 1445.16         
    & 2278.07          
    & \textbf{2585.06}          
    & \textbf{2904.03} 
    & \textbf{2430.84}          
    & 3181.26          
    & \textbf{3964.45}          
    & \textbf{4701.75}          
    & \textbf{2162.83}          
    & \textbf{7465.96}          
    & \textbf{12834.45}          
    & \textbf{23142.83}          
    & \textbf{147193.96}          
    & \textbf{229769.76}          
    & \textbf{402372.32}          
    & \textbf{637621.59}       
    \\\bottomrule
    \end{tabular}
    }
    \caption{Performance comparison under IC diffusion pattern. $-$ indicates out-of-memory error. (Best is highlighted with bold.)}
    \label{tab: evaluation_ic}
    \end{table*}

    \begin{table*}[t]
    \centering
    \resizebox{\textwidth}{!}{%
    \begin{tabular}{@{}c|cccc|cccc|cccc|cccc|cccc|cccc|cccc@{}}
    \toprule
    & \multicolumn{4}{c|}{Cora-ML}
        & \multicolumn{4}{c|}{Celegans}         & \multicolumn{4}{c|}{Arabidopsis}      & \multicolumn{4}{c|}{NYClimateMarch2014}   &  \multicolumn{4}{c|}{ParisAttack2015} 
        \\ \midrule
    Methods & 1\%     & 5\%     & 10\%     & 20\%    & 1\%     & 5\%     & 10\%     & 20\%    & 1\%     & 5\%     & 10\%     & 20\%    & 1\%     & 5\%     & 10\%     & 20\%   & 1\%     & 5\%     & 10\%     & 20\%   
    \\ \midrule
     ISF    & \textbf{381.0}          & \textbf{907.0}          & \textbf{1392.0}          & \textbf{2145.0}       & \textbf{1530.0}           & 2643.0          & \textbf{3274.0}          & 3857.0          & 2901.0          & 4571.0          & 5686.0         & 6855.0          & -          & -          & -          & -          & -          & -           & -          & -           
     \\
    KSN       & 381.0          & 907.0          & 1392.0          & 2145.0   & 1272.0          & 2340.0           & 2959.0         & 3637.0         & 2440.0          & 3969.0          & 4934.0          & 6147.0          & -          & -          & -          & -          & -          & -           & -          & -               \\\midrule
    GCOMB     & 379.0          & 1005.0          & 1297.0          & 2026.0     & 1376.0          & 2327.0          & 3006.0           & 3633.0          & 2501.0           & 4247.0          & 4993.0          & 6286.0          & 5281.0          & 29974.0           & 63187.0          & 74865.0          & 429978.0         & 498301.0          & 571193.0          & 729745.0          
     \\
    ToupleGDD     & 375.0          & 998.0          & 1192.0          & 1989.0        & 1289.0           & 2153.0           & 2932.0          & 3569.0          & 2479.0          & 4147.0          & 4725.0          & 5841.0          & 4866.0          & 26987.0          & 61827.0          & 71865.0          & 420086.0          & 487925.0           & 568872.0          & 739712.0           \\
    DeepIM     & 291.0         & 708.0          & 1023.0          & 1881.0    & 929.0          & 1762.0           & 2012.0          & 2598.0          & 1772.0          & 3232.0          & 3211.0          & 3968.0          & 3582.0          & 21246.0          & 51721.0          & 57977.0          & 387129.0          & 460273.0           & 538160.0          & 578724.0           
    \\
    MIM-Reasoner      & 381.0         & 907.0          & 1392.0          & 2145.0        & 1356.0           & 2340.0          & 3019.0         & 3676.0          & 2554.0          & 4133.0          & 5027.0          & 6191.0          & 6872.0          & 35833.0          & 77925.0          & 96239.0          & 487871.0         & 550864.0           & 629830.0          &  786819.0                \\\midrule
    REM-NonRL        & 321.0          & 811.0          & 1101.0          & 1964.0      & 1043.0           & 2176.0           & 2781.0          & 3246.0          & 2176.0          & 3548.0          & 4185.0          & 5742.0          & 3976.0          & 26920.0          & 55315.0          & 59764.0          & 408149.0          & 468621.0           & 540824.0         & 581872.0        \\
     REM-NonMixture         & 361.0          & 1173.2          & 1212.0          & 2102.0      & 1332.0           & 2347.0           & 3118.0          & 3688.0         & 2632.0          & 4586.0          & 5089.0          & 6197.0          & 5352.0          & 30187.0          & 66872.0          & 77359.0          & 449862.0          & 529562.0           & 592055.0          & 692974.0       \\
    \textbf{REM} & 376.0          & 883.0 & 1281.0           & 2141.0 & 1514.0 & \textbf{2668.0} & 3251.0 & \textbf{3877.0} & \textbf{2943.0} & \textbf{4894.0} & \textbf{5705.0} & \textbf{6890.0} & \textbf{7111.0} & \textbf{37417.0} & \textbf{81255.0} & \textbf{98976.0} & \textbf{503994.0} & \textbf{604890.0} & \textbf{651100.0} & \textbf{804469.0}  \\
    \bottomrule
    \end{tabular}
    }
    \caption{Performance comparison under LT diffusion pattern. $-$ indicates out-of-memory error. (Best is highlighted with bold.)}
    \label{tab: evaluation_lt}
    \end{table*}

\begin{figure*}[!t]
		\subfloat[Cora\_ML-IC]{\label{fig: coraml_ic}
			\hspace{-3mm}\includegraphics[width=0.2\textwidth]{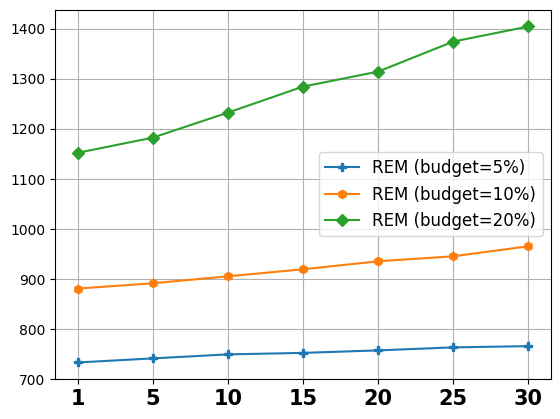}}
		\subfloat[Celegans-IC]{\label{fig: celegans_ic}
			\includegraphics[width=0.2\textwidth]{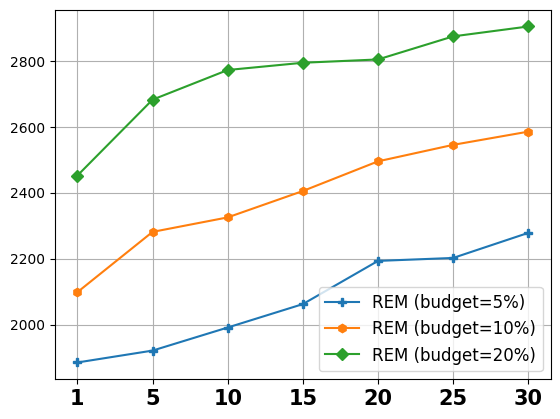}}
		\subfloat[Arabidopsis-IC]{\label{fig: arabidosis_ic}
			\includegraphics[width=0.2\textwidth]{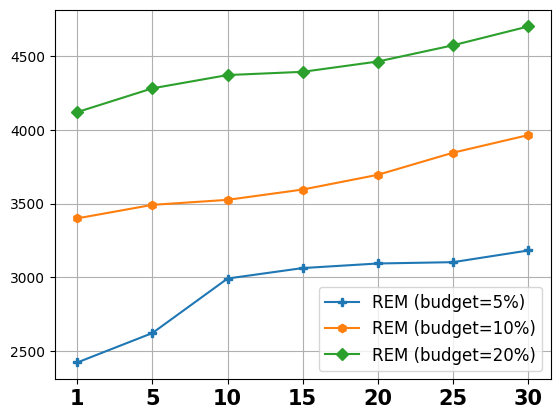}}
		\subfloat[NYClimateMarch-IC]{\label{fig: nyclimate_march_ic}
			\includegraphics[width=0.2\textwidth]{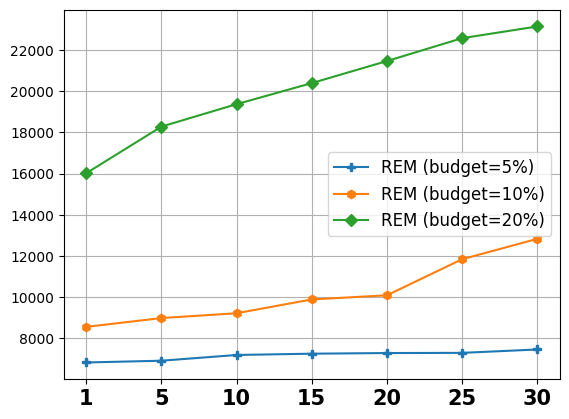}}
		\subfloat[ParisAttack2015-IC]{\label{fig: paris_attack2015_ic}
			\includegraphics[width=0.2\textwidth]{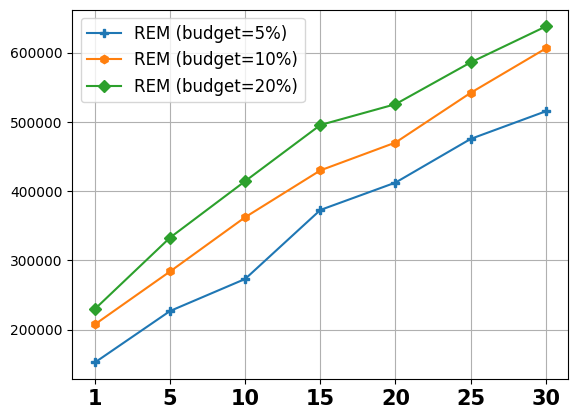}}
		\vspace{-3mm}
		\subfloat[Cora\_ML-LT]{\label{fig: coraml_lt}
			\hspace{-3mm}\includegraphics[width=0.2\textwidth]{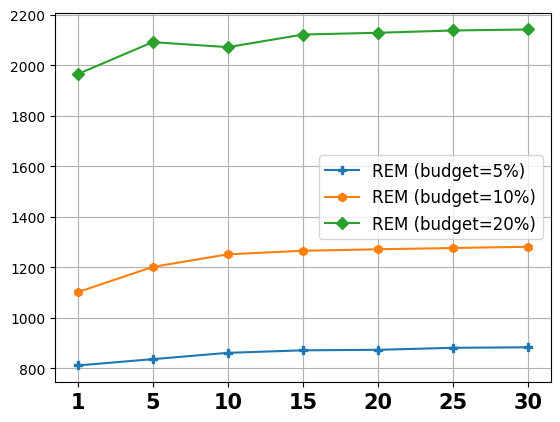}}
		\subfloat[Celegans-LT]{\label{fig: celegans_lt}
			\includegraphics[width=0.2\textwidth]{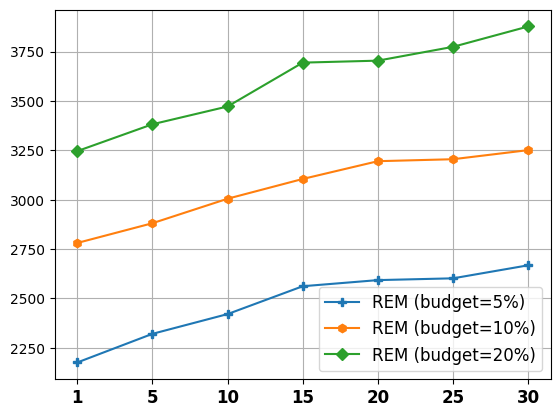}}
		\subfloat[Arabidopsis-LT]{\label{fig: arabidosis_lt}
			\includegraphics[width=0.2\textwidth]{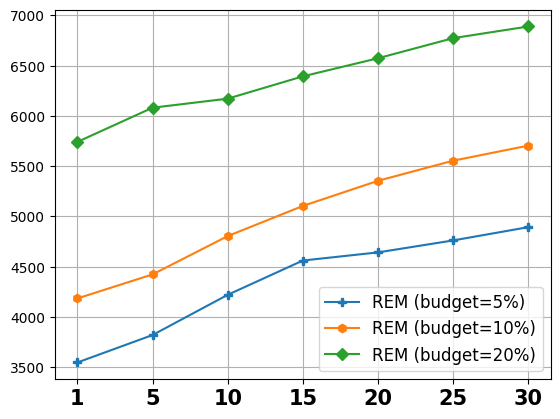}}
		\subfloat[NYClimateMarch-LT]{\label{fig: nyclimate_march_lt}
			\includegraphics[width=0.2\textwidth]{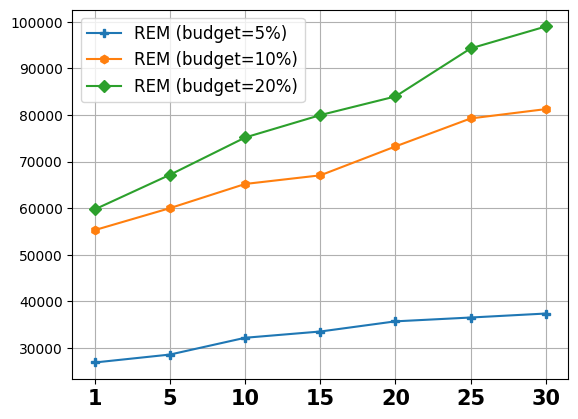}}
		\subfloat[ParisAttack2015-LT]{\label{fig: paris_attack2015_lt}
			\includegraphics[width=0.2\textwidth]{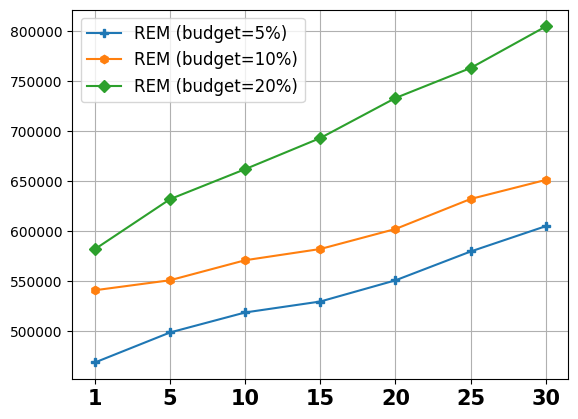}}
			\vspace{-3mm}
		\caption{Difference in influence spread (y-axis) of REM output on different dataset and budget when increasing exploration steps(x axis). Fig. \ref{fig: coraml_ic} - \ref{fig: paris_attack2015_ic} and Fig. \ref{fig: coraml_lt} - \ref{fig: paris_attack2015_lt} are evaluated under the IC and LT model, respectively.}
		\vspace{-5mm}
		\label{fig: budget_constraint}
	\end{figure*}
 
We conduct experiments to compare our proposed REM framework to 6 other state-of-the-art frameworks across 5 real world networks in  various settings.

\subsection{Experiment Setup}

Our main objective is to evaluate the effect of influence spread across different scenarios in Influence Maximization (IM). Our experiments focus on two dominant propagation models within IM: the Linear Threshold (LT) and Independent Cascade (IC) models. To delve deeper into our experimental setup, we refer to Appendix \textbf{D}.

\textbf{Dataset}. Our experiments leverage multiple multiplex network datasets of diverse interaction types and systems. The Celegans Multiplex GPI Network from BioGRID \citep{stark2006biogrid} (version 3.2.108) includes genetic interactions within \textit{Caenorhabditis elegans}, comprising 6 layers, 3,879 nodes, and 8,181 edges. The Arabidopsis Multiplex Network also from BioGRID \citep{stark2006biogrid_arabidopsis} details genetic and protein interactions for \textit{Arabidopsis thaliana}, comprising 7 layers, 6,980 nodes, and 18,654 edges. For social media dynamics, the NYClimateMarch2014 Twitter Network \cite{omodei2015characterizing} captures retweets, mentions, and replies during the People's Climate March, featuring 3 layers, 102,439 nodes, and 353,495 edges. The ParisAttack2015 Twitter Network \cite{dedomenico2020unraveling} includes similar social interactions during the 2015 Paris Attacks, with 3 layers, 1,896,221 nodes, and 4,163,947 edges. We also use the Cora dataset \cite{mccallum2000automating}, a citation network of 2,708 scientific publications and 7,981 edges, to analyze influence in academic publishing.

\subsection{Comparison to other Methods}

We assess the performance of REM by comparing it against two categories of influence maximization techniques.  1) Traditional  methods: \textit{ISF (Influential Seed Finder)} \citep{kuhnle2018multiplex} is a greedy algorithm designed for multiplex influence maximization; \textit{KSN (Knapsack Seeding of Networks)} \citep{kuhnle2018multiplex} utilizes a knapsack approach to find the best seed users in a multiplex network. 2) Deep learning methods: \textit{ToupleGDD} \citep{chen2022touplegdd}, \textit{GCOMB} \cite{manchanda2020gcomb}, \textit{DeepIM} \citep{ling2023deep} are state-of-the-art single network influence maximization solutions.  For multiplex network, the MIM-Reasoner \cite{do2024mimreasoner} method utilize probabilistic graphical models to capture the dynamics within the multiplex, then determine the best seed sets with a reinforcement learning solution. We also evaluate the performance of 2 different REM variants to demonstrate the effectiveness of our approach. One approach is REM-NonRL, which does not employ the exploration of seed sets and solely relies on an initial dataset to provide solution. This variant provides observation on the effectiveness of our proposed reinforcement learning set up. The other variant, REM-NonMixture, forego our Mixture of expert set up, capture the complicated multiplex propagation with one GNN model. This variant will underscore the advantages of our more complex configurations. The comparison is based on three metrics: total influence spread (activated nodes) and inference time (wall-clock time, in seconds).

\subsection{Quantitative Analysis}\label{sec: qua}

We evaluate the performance of the REM method against other IM strategies by comparing their ability to optimize influence across various datasets. In each case, models identify seed nodes representing  1\%, 5\%, 10\%, and 20\% of all nodes. We simulate the diffusion process until completion and determine the average influence spread across 100 iterations. We report the final number infected nodes.

\noindent\textbf{IM under IC Model.} The methods are evaluated on five datasets under the IC diffusion model with budgets of 1\%, 5\%, 10\%, and 20\% of network nodes. As shown in Table \ref{tab: evaluation_ic}, REM consistently outperforms other methods, particularly on large datasets such as NYClimateMarch2014 and ParisAttack2015. Traditional methods (ISF, KSN) perform well on smaller graphs but struggle to scale with larger graphs and higher budgets. Single-graph learning methods (GCOMB, TOUPLEGDD, DEEPIM) fall behind due to their inability to adapt to multiplex networks. While MIM-Reasoner achieves strong results on larger multiplex networks, it is outperformed by REM. Notably, REM variants (REM-NonRL, REM-NonMixture) show significant performance drops, underscoring the importance of REM's key components.

\noindent\textbf{IM under LT Model.} We evaluate the methods under the LT diffusion model, with the results in Table \ref{tab: evaluation_lt} showing that REM consistently outperforms other techniques in maximizing node infections. REM's superiority is particularly evident on large networks and with a $20\%$ seed set, achieving $10\%$ and $15\%$ higher influence spread than the best competing methods on the \textit{NYClimateMarch2014} and \textit{ParisAttack2015} datasets, respectively. This performance highlights REM's superior generalization across diffusion models.

\noindent\textbf{IM with explore step number. } We compare the effectiveness of increasing exploration steps under the IC and LT models within a budget constraint. As shown in Figure \ref{fig: budget_constraint}, more exploration steps generally improve results across networks, especially for larger datasets. For smaller datasets like Cora-ML, the performance difference is minimal, while for larger datasets, the gap widens significantly with more steps.

\subsection{Scalability Analysis}

 \begin{table}[t]
    \centering
    \resizebox{0.9\columnwidth}{!}{%
    \begin{tabular}{@{}l|cccc|c@{}}
\toprule
& \multicolumn{1}{l}{10,000} & 20,000  & 30,000  & 50,000                                        \\ \midrule

GCCOMB & 17.894s  & 30.831s   & 46.275s  &73.983s \\
ToupleGDD & 15.873s  & 25.321   & 37.882s & 58.985s \\
MIM-Reasoner     & \textbf{7.948s }                    & 12.532s  & 26.575s  & 36.437s \\
DeepIM      & 10.321s   & 19.325s   & 32.185s   & 44.871s  
\\ \midrule
\textbf{REM}    & 8.873s   & \textbf{10.198s}   & \textbf{23.404s}   & \textbf{33.482s} \\ \bottomrule
\end{tabular}%
    }
    \vspace{-3mm}
    \caption{The average inference runtime (in seconds) with regard to the increase of node numbers. We select $10\%$ of nodes as the seeds.}
    \label{tab: time}
    \vspace{-7mm}
    \end{table}

We investigate the runtime of seed sets when increasing of graph size of REM verse other learning-based IM solutions. As can be seen in Table \ref{tab: time}, REM demonstrates near-linear growth of runtime as the graph size increases. In addition, it achieves a generally shorter inference time (on average, it has a 10\% faster inference time than the second-fastest MIM-Reasoner and a 20\% faster inference time than the third-fastest DeepIM.

\subsection{Conclusion}

This paper has introduced REM, a framework designed to tackle the inherent challenges of MIM. Through the integration of a Propagtaion Mixture of Experts and a RL-based exploration strategy on a continuous latent representation, REM offers a robust solution to optimize influence spread across multiplex networks. Our approach not only demonstrates superior scalability and efficiency but also excels in handling the diversity of propagation mechanisms within these networks. The empirically experimental results on multiple real-world datasets validate REM’s effectiveness, showcasing its ability to outperform existing state-of-the-art methods in both influence spread and computational efficiency. 

\bibliography{aaai25}

\section{A. Detail Steps of REM}
\label{appendixA}

When solving a MIM problem, REM ultilizes Algorithm \ref{algo: training_framework} for training and Algorithm \ref{algo: prediction} for inference.  Algorithm \ref{algo: training_framework} initializes Seed2Vec and PMoE models, and then iteratively explores the latent space to discover novel seed sets with high propagation potential (lines 6-10). These new sets are combined with the original data to retrain Seed2Vec and PMoE (lines 12-14). Then, Algorithm \ref{algo: prediction} infers the optimal seed set by optimizing a latent representation using gradient ascent to maximize the predicted influence from the PMoE model (lines 3-5).  This representation is then decoded Seed2Vec to output the final solution (line 6). The code and datasets for REM are available at the following GitHub repository: \url{https://github.com/huyenxam/REM}.

\begin{algorithm}[t]
\caption{REM Framework}\label{algo: training_framework}
\begin{algorithmic}[1]
\renewcommand{\algorithmicrequire}{\textbf{Input:}}
\renewcommand{\algorithmicensure}{\textbf{Require:}}
\REQUIRE Multiplex graph $\mathcal{G}$, budget $b$ and the original seed set dataset $\boldsymbol{X}_0$ with $N$ instances.\\
\renewcommand{\algorithmicrequire}{\textbf{Output:}}
\renewcommand{\algorithmicensure}{\textbf{Require:}}
\REQUIRE Optimal seed set $\tilde{\boldsymbol{x}}$.\\
\STATE Initialize Seed2Vec model $\mathcal{F}_\theta$ with encoder $\mathcal{E}_\psi$ and decoder $\mathcal{D}_\phi$.
\STATE Initialize PMoE model $\mathcal{P}(\boldsymbol{x}, \mathcal{G} ; \xi)$.
\STATE Initialize the number of step $T$ and the number of iteration $\eta$, learning rate $\alpha$.
\STATE Initialize Priority Replay Memory (PRM).
\FOR {$t=0, ..., T$}
    \STATE $\boldsymbol{z} \sim \mathcal{N}(\boldsymbol{\mu}, \boldsymbol{\Sigma})$
    \FOR {$i=0, ..., \eta$}
        \STATE $z \leftarrow z - \alpha\cdot \nabla \mathcal{L}^{\text{Explore}}(\boldsymbol{z})$.
        \STATE Reconstruct seed set: $\hat{\boldsymbol{x}} = \mathcal{D}_\phi(\boldsymbol{z})$.
        \STATE Store $(\hat{\boldsymbol{x}}, \mathcal{P}(\mathcal{D}_\phi(\boldsymbol{z}), \mathcal{G}; \xi))$ in PRM.
    \ENDFOR\\
    \STATE Sample top $k$ seed sets $\mathcal{S}_t^{(<k)}$ from PRM based on $\mathcal{P}$. \\
    \STATE Combined dataset $\boldsymbol{X}_t = \mathcal{S}_t^{(<k)} \cup \boldsymbol{X}_0$ \\
    \STATE Retrain $\mathcal{F}_\theta$ and $\mathcal{P}$ on combined dataset. \\
\ENDFOR\\
\STATE $\tilde{\boldsymbol{x}}$ := REM\_PREDICTION$(\phi, \xi, \mathcal{G})$
\STATE \textbf{Return: } $\tilde{\boldsymbol{x}}$
\end{algorithmic}
\end{algorithm}

\begin{algorithm}[t]
\caption{REM\_PREDICTION}\label{algo: prediction}
\begin{algorithmic}[1]
\renewcommand{\algorithmicrequire}{\textbf{Input:}}
\renewcommand{\algorithmicensure}{\textbf{Require:}}
\REQUIRE Decoder parameters $\phi$, PMoE parameters $\xi$, Multiplex graph $\mathcal{G}$.\\
\renewcommand{\algorithmicrequire}{\textbf{Output:}}
\renewcommand{\algorithmicensure}{\textbf{Require:}}
\REQUIRE Optimal seed set $\tilde{\boldsymbol{x}}$.\\
\STATE Initialize the number of iteration $\eta$, learning rate $\beta$.
\STATE Initialize random latent representation $\boldsymbol{z} \sim \mathcal{N}(\boldsymbol{\mu}, \boldsymbol{\Sigma})$.
\FOR {$i=0, ..., \eta$}
\STATE $\boldsymbol{z} \leftarrow \boldsymbol{z} + \beta\cdot \nabla_{\boldsymbol{z}} \mathcal{P}(\mathcal{D}_\phi(\boldsymbol{z}), \mathcal{G}; \xi)$
\ENDFOR\\
\STATE $\tilde{\boldsymbol{x}} = \mathcal{D}_\phi(\boldsymbol{z})$ \\
\STATE \textbf{Return: } $\tilde{\boldsymbol{x}}$
\end{algorithmic}
\end{algorithm}

\section{B. Detail ELBO}
\label{appendixB}

In the Variational Autoencoder (VAE) framework, the primary objective is to maximize the Evidence Lower Bound (ELBO), which serves as a proxy for the log-likelihood of the data. The ELBO comprises two key components: the Reconstruction Loss and the KL Divergence.

\begin{equation}\label{eq: dfn_elbo}
\begin{aligned}
\mathcal{L}^{\text {ELBO }} & =\mathbb{E}_{q_\psi}\left[\log p_\phi(\boldsymbol{x} \mid \boldsymbol{z})\right]-\mathbb{E}_{q_\psi}\left[\log \frac{q_\psi(\boldsymbol{z} \mid \boldsymbol{x})}{p_\phi(\boldsymbol{z})}\right]
\end{aligned}
\end{equation}

The Reconstruction Loss measures the dissimilarity between the original seed set $\boldsymbol{x}$ and its reconstruction $\hat{\boldsymbol{x}}$, while the KL Divergence regularizes the latent space distribution $q_\psi(\boldsymbol{z} \mid \boldsymbol{x})$ towards a prior distribution $p_\phi(\boldsymbol{z})$. The process of minimizing Reconstruction Loss is the role of the Decoder model parameterized by \( \phi \). Specifically, the Decoder observes \( \boldsymbol{z} \) and attempts to generate a reconstruction \( \hat{\boldsymbol{x}} \) that is as close as possible to the original data, thus minimizing the Reconstruction Loss.
To effectively train the VAE, the Mean Squared Error (MSE) loss is used as reconstruction loss. The MSE loss directly quantifies the difference between the original input \( \boldsymbol{x} \) and the reconstructed output \( \hat{\boldsymbol{x}} \), making it a straightforward and widely used loss function for this purpose. The MSE loss is given by:
\begin{equation}
\begin{aligned}
\text{MSE Loss} = \frac{1}{N} \sum_{i=1}^N \left\|\hat{\boldsymbol{x}}_i-\boldsymbol{x}_i\right\|^2
\end{aligned}
\end{equation}

Minimizing the MSE loss corresponds to maximizing the likelihood term \( \log p_\phi(\boldsymbol{x} \mid \boldsymbol{z}) \) within the ELBO. This is because a smaller MSE indicates that the reconstructed output \( \hat{\boldsymbol{x}} \) is closer to the original input \( \boldsymbol{x} \), implying a higher probability of the data under the model.

Finally, the ELBO can be expressed as:
\begin{equation}
\begin{aligned}
\mathcal{L}^{\mathrm{ELBO}}(\boldsymbol{x} ; \psi, \phi)= &\frac{1}{N} \sum_{i=1}^N MSE(\hat{\boldsymbol{x}}_i,\boldsymbol{x}_i)\\
&-\mathrm{KL}\left(q_\psi(\boldsymbol{z} \mid \boldsymbol{x}) \| p_\phi(\boldsymbol{z})\right)
\end{aligned}
\end{equation}

Here, the MSE loss is directly incorporated as the reconstruction term in the ELBO, guiding the optimization process to improve the VAE's ability to reconstruct the input data accurately. Meanwhile, the KL Divergence term ensures that the latent space is regularized towards the prior distribution, maintaining a balance between reconstruction quality and latent space regularization.

\section{C. Proofs}
\label{appendixC}

\subsection{C1. Monotonicity of PMoE Models}
\label{appendixC1}

\textbf{Lemma 1 (Monotonicity of PMoE Models)}
\label{lemma: Monotonicity}
\textit{Assuming the PMoE model has been trained to convergence and during the inference phase, noisy scores $\xi_n$ are not considered, for any GNN-based, $\mathcal{P}$ is infection monotonic if the aggregation function and combine function in GNN are non-decreasing. }

\begin{proof} 

Assuming we have a Graph Neural Network (GNN) with $H$ layers, where $\mathcal{A}^h$ and $C^h$ are  non-decreasing, denoted as $e(.)$. The input is a vector $\boldsymbol{x}$, and we apply the GNN to $\boldsymbol{x}$ over $H$ layers as follows:

1. \textbf{Input Definition}: Initially, consider the input $r_v^{(0)} $ to be $ \boldsymbol{x} $ for every node $ v $ in the graph, meaning all nodes start with the initial feature vector $ \boldsymbol{x} $.

2. \textbf{Iterating Through Layers}: For each layer $ h $ from 1 to $ H $, the aggregation  function is applied to each node $ v $ as follows:

\begin{equation}
e(\boldsymbol{x})=\mathcal{A}^1 \circ\left(C^1 \circ \mathcal{A}^2 \circ C^2 \cdots \circ \mathcal{A}^H \circ C^H\right)
\end{equation}

Because $\mathcal{A}^h$ and $C^h$ are  non-decreasing, so is $\mathcal{A}^1\circ C^{1}\cdots\circ \mathcal{A}^H\circ C^H$, which is $e(\boldsymbol{x})$. \textbf{Therefore, we have that $e(\boldsymbol{x})$ is a non-decreasing function}. 

Now, we will prove that propagation synthesized from a Mixture of Experts model is also a non-decreasing function, provided that the model has converged and there is no noise in expert selection. 
Recall the fact that the propagation $\mathcal{M}(\boldsymbol{x}, \mathcal{G}; \xi)$ given any seed set $\boldsymbol{x}$ can be calculated by Eq. 10. In our setting where we only consider non-noisy experts, therefore $\mathcal{M}(\boldsymbol{x}, \mathcal{G}; \xi)$ is independent of $\xi_n$. We can reformulate the output of our model as:

\begin{equation}
\begin{aligned}
Q\left(\boldsymbol{x},  \xi_g, \xi_n\right)&
=\boldsymbol{x} \xi_g+\epsilon \cdot \operatorname{Softplus}\left(\boldsymbol{x} \xi_n\right) \\
& =\boldsymbol{x} \xi_g+ 0  \cdot \operatorname{Softplus}\left(\boldsymbol{x} \xi_n\right) \\
&=\boldsymbol{x} \xi_g
\end{aligned}
\end{equation}

\begin{equation}
\begin{aligned}
\Rightarrow R\left(\boldsymbol{x}, \xi_g, \xi_n \right) &= R\left(\boldsymbol{x}, \xi_g\right)\\
&=\operatorname{Softmax}\left(\operatorname{TopM}\left(Q\left(\boldsymbol{x},  \xi_g\right), m\right)\right)
\end{aligned}
\end{equation}

Therefore, we have:

\begin{equation}
\begin{aligned}
\Rightarrow \mathcal{M}\left(\boldsymbol{x}, \mathcal{G} ; \xi \right)=\sum_{i=1}^C R_i(\boldsymbol{x}, \xi_g) e_i(\xi_i)
\end{aligned}
\end{equation}

Because $R(\cdot)$ is the softmax operator, which is non-decreasing, and $e(\cdot)$ is also a non-decreasing function, it follows that the PMoE model $\mathcal{M}(\boldsymbol{x}, \mathcal{G}; \xi)$ is monotonic. Consequently, since the function $g(\cdot)$ is non-decreasing as well, the influence propagation function $\mathcal{P}(\boldsymbol{x}, \mathcal{G}; \xi)$ is likewise monotonic.
\end{proof}

\subsection{C2. Latent Entropy Maximization Equivalence}
\label{appendixC2}

\textbf{Lemma 2 (Latent Entropy Maximization Equivalence)}
\textit{Assuming the Seed2Vec model has convergened, we have $\arg \max _{\boldsymbol{z}} \mathcal{H}(\mathcal{D}_\phi(\boldsymbol{z}))
\propto \arg \max _{\boldsymbol{x}} \mathcal{H}(\boldsymbol{x})
$. }

\begin{proof} 
The entropy of a random variable $\boldsymbol{x}$ is given by:

\begin{equation}
\mathcal{H}(\boldsymbol{x})=-\sum_{i=1}^{|\boldsymbol{x}|} p\left(\boldsymbol{x}_i\right) \log p\left(\boldsymbol{x}_i\right)
\end{equation}

Similarly, the entropy of the latent variable $\mathcal{D}_\phi(\boldsymbol{z})$ is:

\begin{equation}
\mathcal{H}\left(\mathcal{D}_\phi(\boldsymbol{z})\right)=-\sum_{i=1}^{|\hat{\boldsymbol{x}}|} p\left(\hat{\boldsymbol{x}}_i\right) \log p\left(\hat{\boldsymbol{x}}_i\right)
\end{equation}

where $\hat{\boldsymbol{x}} = \mathcal{D}_\phi(\boldsymbol{z})$ represents the data reconstructed from the latent variable $\boldsymbol{z}$.

When $\mathcal{F}_\theta$ has converged, the original data $\boldsymbol{x}$ and the reconstructed data $\hat{\boldsymbol{x}}$ are nearly identical, i.e., $\boldsymbol{x} \approx \hat{\boldsymbol{x}}$. Since $\boldsymbol{x}$ and $\hat{\boldsymbol{x}}$ are nearly the same, their entropies are also nearly the same $\mathcal{H}(\boldsymbol{x}) \approx \mathcal{H}\left(\mathcal{D}_\phi(\boldsymbol{z})\right)$. Therefore, maximizing the entropy of the latent variable $\boldsymbol{z}$ is equivalent to maximizing the entropy of the original data $\boldsymbol{x}$: 

\begin{equation}
\arg \max _{\boldsymbol{z}} \mathcal{H}\left(\mathcal{D}_\phi(\boldsymbol{z})\right) \propto \arg \max _{\boldsymbol{x}} \mathcal{H}(\boldsymbol{x})
\end{equation}

\end{proof}

\subsection{C3. Influence Estimation Consistency}
\label{appendixC3}

\textbf{Theorem 3 (Influence Estimation Consistency)}
\label{lemma: Influence Estimation Consistency}
\textit{Given two distinct seed sets $\boldsymbol{x}^{(i)}$ and $\boldsymbol{x}^{(j)}$, with their corresponding latent representations $\boldsymbol{z}^{(i)}$ and $\boldsymbol{z}^{(j)}$ encoded by a Seed2Vec. 
If the reconstruction error is minimized during the training and $\mathcal{P}(p_{\phi}(\boldsymbol{z}^{(i)}), \mathcal{G}; \xi) > \mathcal{P}(p_{\phi}(\boldsymbol{z}^{(j)}), \mathcal{G}; \xi)$, then it follows that $\mathcal{P}(\boldsymbol{x}^{(i)},
\mathcal{G}; \xi) > \mathcal{P}(\boldsymbol{x}^{(j)}, \mathcal{G}; \xi)$. 
}

\begin{proof}
Given the assumptions from Lemma 2, the PMoE model's influence estimation function $\mathcal{P}$ is monotonic, meaning that for any two $\boldsymbol{x}^{(i)} > \boldsymbol{x}^{(j)}$, then $\mathcal{P}(\boldsymbol{x}^{(i)}, \mathcal{G}; \xi) \geq \mathcal{P}(\boldsymbol{x}^{(j)}, \mathcal{G}; \xi)$. Since the Seed2Vec model minimizes reconstruction error, the latent representations $\boldsymbol{z}^{(i)}$ and $\boldsymbol{z}^{(j)}$ preserve the essential properties of their corresponding original seed sets $\boldsymbol{x}^{(i)}$ and $\boldsymbol{x}^{(j)}$. Therefore, if $\mathcal{P}(p_{\phi}(\boldsymbol{z}^{(i)}), \mathcal{G}; \xi) > \mathcal{P}(p_{\phi}(\boldsymbol{z}^{(j)}), \mathcal{G}; \xi)$ in the latent space, the same ordering must hold in the original space, leading to $\mathcal{P}(\boldsymbol{x}^{(i)}, \mathcal{G}; \xi) > \mathcal{P}(\boldsymbol{x}^{(j)}, \mathcal{G}; \xi)$. This completes the proof of Theorem 1.
\end{proof}

\begin{table}[t]
    \centering
    \begin{tabularx}{\linewidth}{|X|c|}
    \toprule
    \textbf{Parameter} & \textbf{Value} \\ \midrule
     Learning rate for VAE Model & 0.003 \\
     Learning rate for PMOE Model & 0.001 \\
    Optimizer & Adam \\
    Number of steps per episode  & 400 \\
    Number of episodes  & 30 \\
    Minibatch size  & 256 \\
    Weight KL & 0.55 \\
    Number of experts & 8 \\
    Dropout ratio & 0.2 \\
    Entropy coefficient & 0.1 \\ \bottomrule
    \end{tabularx}
    \vspace{-3mm}
    \caption{Hyperparameters for Seed2Vec, Latent Seed Set Exploration, and Propagation Mixture of Expert.}
    \label{tab:hyperparameters}
    \vspace{-7mm}
\end{table}

\section{D. More Experiment}
\label{appendixD}

\begin{table*}[t]
    \centering
    \resizebox{\textwidth}{!}{%
    \begin{tabular}{@{}c|cccc|cccc|cccc|cccc|cccc|cccc|cccc@{}}
    \toprule
    &\multicolumn{4}{c|}{Cora-ML}
    & \multicolumn{4}{c|}{Celegans}      
    & \multicolumn{4}{c|}{Arabidopsis}   
    & \multicolumn{4}{c|}{NYClimateMarch2014} 
    &  \multicolumn{4}{c|}{ParisAttack2015}  \\ \midrule
    Methods & 1\%     & 5\%     & 10\%     & 20\%    & 1\%     & 5\%     & 10\%     & 20\%    & 1\%     & 5\%     & 10\%     & 20\%    & 1\%     & 5\%     & 10\%     & 20\%   & 1\%     & 5\%     & 10\%     & 20\%  
    \\ \midrule
    ISF            & \textbf{0.14710} & \textbf{0.28753} & \textbf{0.36184} & 0.50538          & \textbf{0.37790} & \textbf{0.59242} & 0.66304          & 0.72680          & 0.34599          & \textbf{0.44994} & 0.55460          & 0.67252          &                 &                 &                 &                 &                 &                 &                 &                 \\
KSN            & 0.14709          & 0.28753          & 0.36156          & 0.50444          & 0.35650          & 0.56105          & 0.60207          & 0.67546          & 0.32704          & 0.42142          & 0.51881          & 0.66495          &                 &                 &                 &                 &                 &                 &                 &                 \\ \hline
GCOMB          & 0.12818          & 0.28287          & 0.36047          & 0.46216          & 0.35824          & 0.48901          & 0.57679          & 0.65754          & 0.33172          & 0.44384          & 0.51892          & 0.65152          & 0.0204          & 0.0706          & 0.1150          & 0.1653          & 0.0605          & 0.0954          & 0.1878          & 0.3100          \\
ToupleGDD      & 0.12888          & 0.26640          & 0.31847          & 0.41826          & 0.32980          & 0.49117          & 0.54595          & 0.62159          & 0.29293          & 0.40922          & 0.49963          & 0.64230          & 0.0178          & 0.0656          & 0.0999          & 0.1740          & 0.0543          & 0.0907          & 0.1768          & 0.2971          \\
DeepIM         & 0.11504          & 0.22408          & 0.30517          & 0.43554          & 0.32886          & 0.39372          & 0.49976          & 0.58044          & 0.28558          & 0.34354          & 0.47684          & 0.58360          & 0.0185          & 0.0626          & 0.0787          & 0.1393          & 0.0443          & 0.0787          & 0.1483          & 0.2533          \\
MIM-Reasoner   & 0.14705          & 0.28730          & 0.36150          & 0.50345          & 0.36927          & 0.56697          & 0.61611          & 0.68194          & 0.34340          & 0.42824          & 0.53430          & 0.66211          & 0.0205          & 0.0721          & 0.1170          & 0.2056          & 0.0684          & 0.1146          & 0.2004          & 0.3207          \\ \hline
REM-NonRL      & 0.11863          & 0.27062          & 0.32515          & 0.42529          & 0.33545          & 0.48594          & 0.54090          & 0.63209          & 0.29747          & 0.34687          & 0.48701          & 0.59033          & 0.0201          & 0.0667          & 0.0836          & 0.1564          & 0.0459          & 0.0805          & 0.1576          & 0.2717          \\
REM-NonMixture & 0.12672          & 0.27194          & 0.34019          & 0.48063          & 0.35774          & 0.53178          & 0.59401          & 0.69484          & 0.33200          & 0.42732          & 0.53186          & 0.66064          & 0.0216          & 0.0695          & 0.0980          & 0.1848          & 0.0660          & 0.1095          & 0.2005          & 0.3195          \\
\textbf{REM}   & 0.12826          & 0.28267          & 0.35637          & \textbf{0.51852} & 0.37256          & 0.58728          & \textbf{0.66642} & \textbf{0.74865} & \textbf{0.34826} & 0.45577          & \textbf{0.56797} & \textbf{0.67360} & \textbf{0.0211} & \textbf{0.0729} & \textbf{0.1253} & \textbf{0.2259} & \textbf{0.0776} & \textbf{0.1212} & \textbf{0.2122} & \textbf{0.3363} \\ \hline
    
    \\\bottomrule
    \end{tabular}
    }
\caption{Comparison of the percentage of infected nodes under the IC diffusion pattern.  "-" indicates out-of-memory error. (Best results are highlighted in bold.)}
    \label{tab: evaluation_ic_percentage}
    \end{table*}

    \begin{table*}[t]
    \centering
    \resizebox{\textwidth}{!}{%
    \begin{tabular}{@{}c|cccc|cccc|cccc|cccc|cccc|cccc|cccc@{}}
    \toprule
    & \multicolumn{4}{c|}{Cora-ML}
        & \multicolumn{4}{c|}{Celegans}         & \multicolumn{4}{c|}{Arabidopsis}      & \multicolumn{4}{c|}{NYClimateMarch2014}   &  \multicolumn{4}{c|}{ParisAttack2015} 
        \\ \midrule
    Methods & 1\%     & 5\%     & 10\%     & 20\%    & 1\%     & 5\%     & 10\%     & 20\%    & 1\%     & 5\%     & 10\%     & 20\%    & 1\%     & 5\%     & 10\%     & 20\%   & 1\%     & 5\%     & 10\%     & 20\%   
    \\ \midrule
     ISF            & \textbf{0.1407} & \textbf{0.3349} & \textbf{0.5140} & \textbf{0.7921} & \textbf{0.3944} & 0.6814          & \textbf{0.8440} & 0.9943          & 0.4156          & 0.6549          & 0.8146          & 0.9821          &                 &                 &                 &                 &                 &                 &                 &                 \\
KSN            & 0.1407          & 0.3349          & 0.5140          & 0.7921          & 0.3279          & 0.6032          & 0.7628          & 0.9376          & 0.3496          & 0.5686          & 0.7069          & 0.8807          &                 &                 &                 &                 &                 &                 &                 &                 \\ \hline
GCOMB          & 0.1400          & 0.3711          & 0.4790          & 0.7482          & 0.3547          & 0.5999          & 0.7749          & 0.9366          & 0.3583          & 0.6085          & 0.7153          & 0.9006          & 0.0516          & 0.2926          & 0.6168          & 0.7308          & 0.2268          & 0.2628          & 0.3012          & 0.3848          \\
ToupleGDD      & 0.1385          & 0.3685          & 0.4402          & 0.7345          & 0.3323          & 0.5550          & 0.7559          & 0.9201          & 0.3552          & 0.5941          & 0.6769          & 0.8368          & 0.0475          & 0.2634          & 0.6035          & 0.7015          & 0.2215          & 0.2573          & 0.3000          & 0.3901          \\
DeepIM         & 0.1075          & 0.2614          & 0.3778          & 0.6946          & 0.2395          & 0.4542          & 0.5187          & 0.6698          & 0.2539          & 0.4630          & 0.4600          & 0.5685          & 0.0350          & 0.2074          & 0.5049          & 0.5660          & 0.2042          & 0.2427          & 0.2838          & 0.3052          \\
MIM-Reasoner   & 0.1407          & 0.3349          & 0.5140          & 0.7921          & 0.3496          & 0.6032          & 0.7783          & 0.9477          & 0.3659          & 0.5921          & 0.7202          & 0.8870          & 0.0671          & 0.3498          & 0.7607          & 0.9395          & 0.2573          & 0.2905          & 0.3322          & 0.4149          \\ \hline
REM-NonRL      & 0.1185          & 0.2995          & 0.4066          & 0.7253          & 0.2689          & 0.5610          & 0.7169          & 0.8368          & 0.3117          & 0.5083          & 0.5996          & 0.8226          & 0.0388          & 0.2628          & 0.5400          & 0.5834          & 0.2152          & 0.2471          & 0.2852          & 0.3069          \\
REM-NonMixture & 0.1333          & 0.4332          & 0.4476          & 0.7762          & 0.3434          & 0.6051          & 0.8038          & 0.9508          & 0.3771          & 0.6570          & 0.7291          & 0.8878          & 0.0522          & 0.2947          & 0.6528          & 0.7552          & 0.2372          & 0.2793          & 0.3122          & 0.3655          \\
\textbf{REM}   & 0.1388          & 0.3261          & 0.4730          & 0.7906          & 0.3903          & \textbf{0.6878} & 0.8381          & \textbf{0.9995} & \textbf{0.4216} & \textbf{0.7011} & \textbf{0.8173} & \textbf{0.9871} & \textbf{0.0694} & \textbf{0.3653} & \textbf{0.7932} & \textbf{0.9662} & \textbf{0.2658} & \textbf{0.3190} & \textbf{0.3434} & \textbf{0.4242} \\
    \bottomrule
    \end{tabular}
    }
\caption{Comparison of the percentage of infected nodes under the LT diffusion pattern.  "-" indicates out-of-memory error. (Best results are highlighted in bold.)}
    \label{tab: evaluation_lt_percentage}
    \end{table*}
    
\subsection{D1. Hyperparameter Setting.}
\label{appendixD1}

We conducted our experiments on a system equipped with an Intel(R) Core i9-13900k processor, 128 GB RAM, and two Nvidia RTX 4090 GPUs with 24GB VRAM each. For each baseline, we set hyperparameters according to their original papers and  fine-tune them on each dataset. For the configuration of each diffusion model, we use a weighted cascade version of the IC model, i.e., the propagation probability $p_{u,v}=1/d^{in}_v$ ($d^{in}_v$ denotes the in-degree of node $v$) for each edge $e=(u, v)$ on graph $G$; For the LT model, the threshold $\theta$ was set to $0.5$ for each node $v$. 

This section outlines the hyperparameter selection for REM (Table \ref{tab:hyperparameters}), focusing on model performance, training stability, and efficiency. Learning rates were adjusted to model complexity: 0.003 for the Variational Autoencoder (VAE) to accelerate convergence, and 0.001 for the over-parameterized Propagation Mixture of Experts (PMOE) to ensure stability. The VAE's KL divergence weight is set to 0.55, balancing reconstruction accuracy and latent space regularization. We specify the number of experts in PMOE to 20, to ensure comprehensive capture of inter-layer processes while mitigating the risk of overlooking potential propagation pathways. A minibatch size of 256 balances training stability of both models.
We configured the latent seed set exploration process for 30 episodes, each comprising 400 steps, resulting in a total of 12,000 generated data points. This volume is sufficient to augment any of the graph architectures under study. Proximal Policy Optimization (PPO) was employed for policy training, incorporating an entropy coefficient and dropout to encourage exploration and prevent overfitting, respectively.

\subsection{D2. Case Study: Graph Neural Network}
\label{appendixD2}
\begin{table}[t]
    \centering
    \resizebox{1.\columnwidth}{!}{%
    \begin{tabular}{|@{}c|cc|cc|cc|cc@{}}
    \toprule
    & \multicolumn{2}{c|}{\textbf{IC}}
        & \multicolumn{2}{c|}{\textbf{LT}}        
        \\ \midrule
Methods & GCN     & GAT     & GCN     & GAT  \\ \midrule
Cora-ML & 947.3  & \textbf{965.04} & 1235.0  & \textbf{1281.0}   \\
Celegans & 2419.2  & \textbf{2585.06} & 3182.0  & \textbf{3251.0}   \\
Arabidopsis     & 3802.4  & \textbf{3964.45} & 5398.0   & \textbf{5705.0} \\
NYClimateMarch2014      & 11198.3  & \textbf{12834.45} & 76912.0   & \textbf{81255.0} \\ 
ParisAttack2015    & 397672.1  & \textbf{402372.32} & 603127.0   & \textbf{651100.0}  \\ \bottomrule
\end{tabular}%
}
\vspace{-3mm}
\caption{Comparison between choosing GCN and GAT as expert architecture in PMoE, with $10\%$ of nodes as budget, under both LT and IC.}
\label{tab: gnn_based}
\vspace{-7mm}
\end{table}
This section compares the performance between two prominent Graph Neural Network (GNN) architectures, Graph Convolutional Network (GCN) and Graph Attention Network (GAT), within our PMoE framework. Table \ref{tab: gnn_based} presents the influence spread achieved by REM when applying each GNN variant as its PMoE expert architecture on the 5 aforementioned datasets, under both LT and IC diffusion models. Notably, we observe superior performance with GAT compared to GCN across all scenarios. This difference in performance arises from GAT's ability to assign varying levels of importance to neighboring nodes during the aggregation process, unlike GCN, which treats all neighbors equally. Choosing GAT architecture for the experts is neccessary, especially in multiplex, where the diverse and complex nature of node relationships demands a more adaptive and selective aggregation process to achieve optimal performance.

\subsection{D3. Final infected nodes percentage}
\label{appendixD3}

In addition to our final total infected node results, we added the percentage of nodes infected in the graph. This percentage is computed by dividing the number of nodes activated by the end of the diffusion process by the total number of nodes in the network. This metric provides a direct comparison of the effectiveness of different IM strategies in terms of their relative reach within the network.

\noindent\textbf{IM under IC Model.} 
Table \ref{tab: evaluation_ic_percentage} compares the percentage of infected nodes (infected nodes / total nodes) achieved by various IM methods on five datasets under the IC diffusion model with four seed set budgets (1\%, 5\%, 10\%, 20\%). REM consistently outperforms others, especially on larger datasets (NYClimateMarch2014, ParisAttack2015). ISF and KSN perform well on small graphs but lack scalability. Single-network methods (GCOMB, ToupleGDD, DeepIM) struggle on multiplex networks. MIM-Reasoner performs well but is surpassed by REM, with REM-NonRL and REM-NonMixture variants underscoring the value of reinforcement learning and Mixture of Experts in REM.

\noindent\textbf{IM under LT Model.} 
Table \ref{tab: evaluation_lt_percentage} shows the percentage of infected nodes for each method under the LT model across the same datasets and seed set budgets. REM achieves the highest percentages in most cases, excelling on larger datasets with larger budgets. For example, at 20\% budget, REM outperforms the next best method on NYClimateMarch2014 and ParisAttack2015 by ~10\% and ~15\% respectively. Results highlight REM's effectiveness in maximizing influence spread under the LT model and the impact of increasing seed set budgets.

\end{document}